\documentclass[preprint,12pt]{elsarticle}

\def\ben{\begin{enumerate}} \def\een{\end{enumerate}}
\def\beq{\begin{equation}} \def\eeq{\end{equation}}
\def\beqn{\begin{equation*}} \def\eeqn{\end{equation*}}
\def\bea{\begin{eqnarray}} \def\eea{\end{eqnarray}}
\def\ba{\begin{array}} \def\ea{\end{array}}
\def\beann{\begin{eqnarray*}} \def\eeann{\end{eqnarray*}}
\def\beasn{\begin{sneqnarray}} \def\eeasn{\end{sneqnarray}}
\def\fn{\footnote}

\usepackage{amssymb}

\begin{document}

\begin{frontmatter}



\title{Leon Rosenfeld and the challenge of the vanishing momentum in quantum electrodynamics}


\author[label1,label2]{Donald Salisbury}
 \ead{dsalisbury@austincollege.edu}
\address[label1]{Department of Physics,
Austin College, Sherman, Texas 75090-4440, USA}

\address[label2]{Max Planck Institut f\"ur Wissenschaftsgeschichte,
Boltzmannstrasse 22,
14195 Berlin, Germany}

\begin{abstract}
Leon Rosenfeld published in 1930 the first systematic Hamiltonian approach to Lagrangian models that possess a local gauge symmetry. The application of this formalism to theories with local internal symmetries, such as electromagnetism in interaction with charged matter fields, is valid and complete, and predates by two decades the work by Dirac and Bergmann. Although he provided a group-theoretical justification for gauge fixing procedures that had just been implemented in the first expositions of quantum electrodynamics by Heisenberg and Pauli, and also by Fermi, his contribution went largely unnoticed. This lack of impact seems to be related to a generalized disenchantment with second quantization in the 1930's and 1940's.

\end{abstract}


\end{frontmatter}

\section{Introduction}

Leon Rosenfeld is best known for his treatment with Niels Bohr in 1933 of the measurability of quantum electrodynamic fields \cite{Bohr:1933aa}\footnote{An English translation appears in \cite{Bohr:1979aa}}. Less well-known is his groundbreaking analysis of the phase space implementation of gauge symmetries that he published in {\it Annalen der Physik} in 1930 under the title ``Zur Quantelung der Wellenfelder" (On the Quantization of Wave Fields) \cite{Rosenfeld:1930aa}\footnote{An English translation of this paper, with extensive annotation, will appear as a Max Planck Institute for the History of Science preprint. A critical analysis will appear in {\it Archive for History of Exact Sciences.}}. In this paper I will discuss Rosenfeld's invention of constrained Hamiltonian dynamics, and in particular his application of this formalism to quantum electrodynamics. Ultimately we would like to know why neither Wolfgang Pauli, who had recommended this analysis to Rosenfeld, nor apparently anyone else in the ensuing twenty years, acknowledged the pertinence of this work for the development of quantum electrodynamics. Indeed, it was only following the work of Peter G. Bergmann and P. A. M. Dirac, commencing in 1949\footnote{Both Dirac\cite{Dirac:1950aa,Dirac:1951aa} and Bergmann and his collaborators\cite{Bergmann:1949aa,Bergmann:1949ab,Bergmann:1950ab,Anderson:1951aa} laid out the general formalism in the period from 1949 to 1951 }, that the systematic treatment of constrained Hamiltonian systems began to attract attention. This formalism is now known as the Dirac-Bergmann procedure.  Bergmann's interest was in the Hamiltonian version of Einstein's general theory of relativity, as a first step in its eventual quantization. He was initially not aware of Rosenfeld's work, but when he did learn of it he consistently cited it as a forerunner of his own work. Dirac on the other hand, as we shall see below, was already in 1932 aware of Rosenfeld's formalism, specifically in regard to it's application to quantum electrodynamics. Yet as far as I can tell Dirac never acknowledged Rosenfeld's contribution. I do not wish to debate priorities in this paper. Rather, we shall attempt to understand the contextual dynamic of this story with the hope that it will shed light not only on the early development of quantum electrodynamics but also on the subsequent development of gauge theories.

Initial progress with canonical electrodynamic quantum field theory was temporally stymied in 1929 by the identical vanishing of the momentum associated with the temporal component of the electromagnetic potential. I will first briefly review the earlier history of quantum electrodynamics, then discuss the not altogether satisfactory resolution of this quandry that was published by Pauli and Werner Heisenberg. This is where Rosenfeld enters the stage. Following a brief biographical sketch I will then review his pioneering constrained dynamics formalism with a description in detail of his application of the program to Lorentz covariant electrodynamics. Then I will address the resounding lack of impact of his contribution. An appendix contains a group theoretical discussion of the canonical imposition of gauge conditions. 

\section{Quantum electrodynamics before 1930}

The understanding of the interaction between electrically charged matter and the electromagnetic field was of course a focus of the emerging theory of quantum mechanics from the very beginning of its history. Abraham Pais \cite{Pais:1982aa} has argued that quantum mechanics and quantum electrodynamics both found their origins in Planck's  black body energy density formula of 1900 \cite{Planck:1900aa}, though the dates and individuals he attributes to these beginnings might surprise.  He would have Einstein in 1906 as the first to have quantized the material oscillator.\fn{\cite{Pais:1982aa} p. 378}  Debye \cite{Debye:1910aa} in 1910 is the first to have quantized the free radiation field  - and he thereby derived the Planck formula. He achieved this result by treating  each oscillation mode as an independent quantized harmonic oscillator. However, there was substantial resistance to the idea of quantizing the radiation field. The 1924 Bohr Kramers Slater program \cite{Bohr:1924aa} actually represents the final failed effort within the framework of the old quantum theory to confine quantum effects to the electromagnetic interaction with charged matter; the electromagnetic field itself was thought to remain subject to the classical Maxwell description. That the photon particle actually existed was then demonstrated irrefutably first by Bothe and Geiger \cite{Bothe:1925aa} and then by Compton and Simon \cite{Compton:1925aa}  in 1925.  The former work already led Bohr to lament  that  ``Under these circumstances we must be prepared for the fact that the generalization of classical electrodynamic theory that we are seeking will require a thoroughgoing revolution of the concepts on which the description of nature has until now been based.''\footnote{``Bei dieser Sachlage muss man darauf vorbereitet sein, dass die zu erstrebende
Verallgemeinerung der klassischen elektrodynamischen Theorie
eine durchgreifende Revolution der Begriffe fordert, auf denen die Naturbeschreibung
bis jetzt beruht hat.'' \cite{Bohr:1925aa}, p. 155} 

This revolution followed in very short order, in step with the emergence of the new quantum theory. Indeed, its creators almost invariably sought to   broaden the scope of new technical and conceptual insights to include electromagnetic interactions. Often they did this in their original groundbreaking papers. So, for example, in  1925 Born and Jordan \cite{Born:1925aa}, following Heisenberg's lead, proposed that the electric and magnetic fields ought to be represented by matrices. This suggestion preceded the epochal deduction of the position and momentum commutation relation for finite dimensional systems in their joint paper the following year with Heisenberg \cite{Born:1926aa}, and they did not inquire into the electromagnetic field algebra.  Jordan's contribution in the latter paper is often cited as the beginning of quantum field theory. He introduced canonical commutation relations for the Fourier modes of a field theory with one spatial dimension: a string. He was able to calculate the mean squared  energy fluctuations for this theory, obtaining a sum of two expressions - one of which was clearly of particle origin and the other clearly the result of wave interference. Such an expression had originally been obtained by Einstein \cite{Einstein:1909aa} from Planck's energy density formula by applying statistical mechanical arguments. We witness here an instance in which the obtainment of a desired quantum field theoretical result actually buttressed the belief of researchers that they were making progress in formulating a correct theory of quantum mechanics for finite systems.

It was Dirac who first saw the relation between the commutation rule $q p - p q = i \hbar$ and Poisson brackets, thus creating a general algebraic canonical quantization rule, a rule that did not necessitate the use of matrices \cite{Dirac:1925aa}. He recounts in his 1977 Varenna lectures that the idea came to him ``in a flash".\footnote{\cite{Dirac:1977aa}, p. 122} Hamiltonian dynamics was not at that time a staple in the education of a young physicist, but he had had already used it extensively. The classical canonical transformation formalism suggested to him a quantum mechanical analogue that he dubbed ``transformation  theory''. In modern parlance one important aspect of Dirac's theory is that it offers a freedom to change representations. It would therefore serve as a basis for the demonstration of the equivalence of the Heisenberg-Born-Jordan matrix representation and Schr\"odinger's wave mechanics, to be addressed shortly. But perhaps the most important aspect of Dirac's transformation theory for this essay is that it provides a means for translating classical canonically implemented symmetries as transformations of quantum variables. Strangely, in a 1972 historical talk on the occasion of Dirac's seventieth birthday, Res Jost \cite{Jost:1972aa} also notes that ``Dirac's deep affinity for analytical dynamics is still noticeable in his papers on quantum electrodynamics...''. But then he goes on to remark that ``... this is a use that we would hardly find justified in our own time''. This is consistent with the idea that widespread interest in canonically implemented gauge symmetries really grew only after t'Hooft's proof in 1971 \cite{Hooft:1971aa} of the renormalizability of non-Abelian gauge theories. The thesis is supported in the same volume by C. Lanczos \cite{Lanczos:1972aa}, who notes that almost no one involved in the development of quantum mechanics at this time recognized the group-theoretical significance of Dirac's quantum transformation theory.  

As he recounted in his Varenna lectures, Dirac's dominant interest early in his graduate career was in relativity theory. Combined with his expertise in Hamiltonian dynamics it was natural for him to work out independently a Hamiltonian formulation of charged particles in interaction with an external electromagnetic field. He writes, ``When I first met this problem, I proceeded to solve it without bothering to look up the literature to see whether it had been solved previously... [It] did not involve much difficulty, and I think it was much simpler than looking up the references.''\footnote{\cite{Dirac:1977aa}, p. 143}. I quote this in part to highlight a recurrent feature of Dirac's work, his tendency to follow up on hunches without undertaking extensive literature searches. Dirac's first foray into relativistic Hamiltonian dynamics resulted in a remarkable treatment of the Compton effect, employing action and angle variables in the context of the old quantum theory \cite{Dirac:1926ab}. The work is innovative in two respects. First, he managed to get the desired result while still treating the radiation field as a classical external field. We find here also very likely the first appearance of a quantum Hamiltonian constraint: he promotes the time to a quantum operator with the consequence that Hamiltonian vanishes. It was this work that occasioned  the following interchange with Thomas Kuhn in a 1963 {\it Archive for the History of Science} Interview \footnote{AHQM, May 10, 1963, p 15}
\begin{quote} Kuhn: You develop it classically first and then simply apply commutation relations to W and t; the classical formulation is one that I hadn't seen  ...

Dirac: I think it is rather standard that you can count time as an extra variable and introduce something conjugate to it.

Kuhn: Do you think it was relatively standard at the time? I don't know of another place where this point had been put previously in this way, but I'm not at all sure it hadn't.

Dirac:  Well, I think I might answer you in much the same way that I wrote that I felt it had probably been done before, but it was less trouble to me to present it as something new than to search for a reference. A good deal of my work was like that. It happened rather often that there was something which I thought had been done before, but it seemed a great nuisance to look through all the references to try to find it, and if it doesnÕt take much trouble to publish it, one can publish it again without claiming either that it is new or that it has been done before. For example the Hamiltonian theory for an electron moving in a given external field. I think that's probably been done before although I haven't found the reference to it. Just  the classical equation, working with the Hamiltonian which is quadratic in delta as well as quadratic in the three momentum variables. \end{quote}

Regarding the interaction of charged particles with electromagnetic fields, Minkowski in 1908 was the first to write down the manifestly Lorentz covariant ponderomotive equations. I do not know who was the first to publish the Hamiltonian for charged particles in interaction with an external electromagnetic field. It is curious that some of the first efforts in deriving an action for a closed system of interacting charged particles and fields were undertaken in five dimensional models. Schr\"odinger \footnote{\cite{Schrodinger:1927aa}, p. 265} noted that his formulation of a Lagrangian model of matter fields in interaction with an electromagnetic field was a result that was implicit in unified five-dimensional models of Oskar Klein \cite{Klein:1926aa} and of de Donder \cite{Donder:1926aa} in 1926.

\subsection{Pauli, Jordan and Heisenberg}

\subsubsection{Early chronology}

Heisenberg indicates to Pauli in an extensive letter written from Copenhagen in February, 1927, that he is in agreement with Pauli's suggestion that the electromagnetic field be quantized. But then he suggests that ``one should perhaps later also quantize the de Broglie [matter] waves.''\footnote{``Aber man soll dann vielleicht doch auch sp\"ater die de Broglie Wellen quanteln..." \cite{Hermann:1979aa}, p. 376} Pauli indicates over the next three months in letters to Heisenberg\footnote{\cite{Hermann:1979aa}, p. 382}, Jordan and Wentzel that he is busy studying Volterra's functional methods with the intention of applying the canonical Hamiltonian procedure to fields. To Jordan he writes from Hamburg on March 12 that ``we'll see if I'm able to wangle quantum electrodynamics. At the moment I am of good cheer.''\footnote{``Wir werden ja sehen, ob ich die Quantenelecktrodynamik zustande bringe. Vorl\"aufig bin ich guten Mutes.'' \cite{Hermann:1979aa}, p. 386} He continues to express optimism in a letter to Bohr on March 
29. ``I am at the moment very much occupied with quantum electrodynamics... It almost seems as though the solution to the quantum puzzle is not far off..."\footnote{``Momentan bin ich sehr mit Quantenelektrodynamik besch\"aftigt... Es hat fast den Anschein, als ob die L\"osung des Quantenr\"atsels nicht mehr sehr fern liegt.'' \cite{Hermann:1979aa}, p. 389}. But on May 16 he writes to Wentzel that ``there are still unresolved difficulties, and I still have nothing final.''\footnote{``Aber hierbei sind noch ungel\"oste Schwierigkeiten geblieben und ich habe noch nichts Fertiges dar\"uber'' \cite{Hermann:1979aa}, p. 393}

Heisenberg writes Pauli on the same date that he (Heisenberg) ...`` still has no definitive opinion on the Hamiltonian for representing interaction between radiation and matter, I must first study Dirac.''\footnote{``\"Uber die Hamiltonsche Wechselwirkungsfunktion zwischen Strahlung und Materie hab' ich noch keine endg\"ultige Meinung, ich muss Dirac erst studieren.''  \cite{Hermann:1979aa}, p. 395 } Heisenberg described these misgivings in an interview with T. Kuhn and J. L. Heilbron in 1963. His principle concern was with Lorentz covariance. ``As soon as I thought about this side of the problem, then I realized that I was not satisfied with Dirac's papers because then I felt that the introduction of a Coulomb field, besides the light quanta, was ruining the Lorentz invariance from the beginning and then I felt it was very difficult to get it backwards then from the end.''\footnote{Archive for the History of Science, Interview with Heisenberg, July 12, 1963, 74-2}

Still in June, Pauli writes Wentzel: ``Concerning the possibility of proceeding further with quantum electrodynamics I am once again very skeptical.''\footnote{``Hinsichtlich der M\"oglichkeit, mit der Quantenelectrodynamik weiterzukommen, bin ich wieder sehr skeptisch.''  \cite{Hermann:1979aa}, p. 399}  Meanwhile, Pauli teams with Jordan to produce a manifestly Lorentz covariant quantization of the pure electromagnetic radiation field.\cite{Jordan:1928ab}  The gauge invariant electric and magnetic fields are introduced directly as operators; the electromagnetic potentials are not required. Pauli trumpets their success in June in a letter to Bohr: ``Jordan and I have succeeded in formulating these conditions so that, as opposed to Dirac where the position coordinates are distinguished from the time, all of the demands of relativistic invariance are fulfilled.'' He continues ``The pure electrodynamics of the charge-free space can naturally not offer the possibility of measuring field strengths. This will only be possible with electrodynamic observations with particles. But nevertheless we have created a general framework that will be appropriate for their inclusion. I will not be able to think about the relativistic treatment of the many-body problem until after the vacation. Meanwhile neither Jordan or I know anything definite about it. Nevertheless I have some hope.''\footnote{``Jordan und mir ist es gelungen, diese Bedingungen so zu formulieren, dass nicht mehr, wie bei Dirac, die Raum-Koordinaten vor der Zeit ausgezeichnet, sondern vielmehr alle Forderungen der relativistischen Invarianz erf\"ullt sind.'' ``Die blosse Elektrodynamik des ladungsfreien Raumes kann nat\"urlich noch keine Messungsm\"oglichkeiten der Feldst\"arken wirklich angeben. Dies wird erst durch Betrachtungen Ÿber die Elektrodynamik  mit Teilchen m\"oglich sein. Aber immerhin wird f\"ur diese doch bereits ein allgemeiner Rahmen geschaffen, in das sie hineinpassen muss. Zum Nachdenken Ÿber die relativistische Behandlung des Mehrk\"orpersproblems werde ich aber erst nach Ferien kommen, vorl\"aufig wissen weder Jordan noch ich etwas bestimmtes dar\"uber. Immerhin habe ich doch gewisse Hoffnungen.'' \cite{Hermann:1979aa}, pp. 403, 404}

But the interaction problem persists. In 1928 Pauli accepts a position at the ETH in Zurich. In June he reports to Bohr, ``I have hardly made any progress with quantum electrodynamics. Neither has Heisenberg. The difficulties that I described to you in my visit to Copenhagen seem to be very deep and I think that they will be overcome only with a basically new idea. (Was Dirac in Copenhagen? What does he think about this current situation?)''\footnote{``In der Quantenelektrodynamik bin ich gar nicht mehr vorw\"arts gekommen (Heisenberg \"ubrigens auch nicht). Die Schwierigkeiten von denen ich bei meinem Besuch in Kopengagen erz\"ahlte, scheinen doch von sehr tiefliegender Art zu sein und ich glaube jetzt, dass sie erst durch eine prinzipiell neue Idee umgangen werden k\"onnen. (War Dirac in Kopenhagen? Was meint er zur jetztigen Situation?)'' \cite{Hermann:1979aa}, p. 462} The problem begins to appear insurmountable to Pauli. Finally, six months later he writes to Bohr, admitting the depth of his despair. ``Unfortunately nothing is happening with my own work  ... I have been so ashamed about this that I have been putting off writing to you... IÕm just stupid and lazy... Heisenberg has an idea how our relativistic many-body model can perhaps be worked out.''\footnote{``Mit meinen eigenen Arbeiten geht es leider vorl\"aufig gar nicht ... Dar\"uber sch\"amte ich mich so, dass ich es immer hinaus- geschoben habe, Dir zu schreiben... Ich bin nur dumm und faul... Heisenberg hat eine Idee, wie unser Ansatz zum relativistischen Mehrk\"orperproblem vielleicht doch durchgef\"uhrt werden k\"onnte. '' \cite{Hermann:1979aa}, p. 485} In a letter to O. Klein a month later, on February 18, 1928, he reveals that the crisis had led to a suspension of his research. ``For my own amusement I put together at that time a short outline for a utopian novel, with the intended title GulliverÕs Travels to Uranien. It was conceived in Swift's style as a political satire on contemporary democracy, against all that only remotely smacks of parliaments, elections and majorities. Caught up in such dreams there came to me suddenly in January news from Heisenberg (to be explained in the following). He had found a trick to remove the formal difficulties that had stood in the way of our completion of quantum electrodynamics. Our relativisitic many-body problem is practically solved!''\footnote{``Zu meinem eigenen Amusement machte ich damals einen kurzen Entwurf zu einem utopischen Roman, der den Titel Gullivers Reise nach Uranien haben sollte und im Stile vom Swift als politische Satire gegen die heute Demokratie gedacht war, n\"amlich gegen alles, was auch nur entfernt nach Parlamenten, Abstimmungen und Majorit\"aten riecht! In Solchen Tr\"aumen befangen, kam mir im Januar pl\"otzlich eine Nachricht von Heisenberg zu, dass er mittels eines (im Folgenden n\"aher zu erl\"autern) Kunstgriffes die formalen Schwierigkeiten beseitigen kann, die der Durchf\"uhrung unserer Quantenelektrodynamik entgegen standen, so dass das relativistische Mehrk\"orperproblem jetzt gewissermassen  gel\"ost ist!''\cite{Hermann:1979aa}, p. 488}

\subsection{Heisenberg, Pauli and Quantum Electrodynamics}

The problem that had stumped Heisenberg and Pauli was the following. In classical electrodynamics  the momentum $p^0$ conjugate to the temporal component of the electromagnetic potential, $A_0$, vanishes identically; the time derivative $\dot A_0$ does not appear in the pure electromagnetic radiation Lagrangian nor in the interaction terms. Consequently there appeared to be a contradiction with the newly discovered quantization rules. A vanishing $p^0(\vec x)$ operator is inconsistent with the commutator $[A_0(\vec x), p^0(\vec y) ] = i \hbar \delta^3(\vec x, \vec y )$. This problem could be avoided in vacuum electrodynamics since the scalar potential could simply be eliminated from the quantum theory. Or as Pauli and Jordan had shown, one could work exclusively with electric and magnetic field operators. Although they did not reveal this connection in their paper, their relativistically covariant commutation relations are precisely those that obtain through a canonical quantization of the transverse vector potential $\vec A_t$, satisfying $\vec \nabla  \cdot \vec A_t = 0$. Thus they effectively eliminated both the longitudinal $\vec A$ field and the scalar field $A_0$. Dirac,  in his groundbreaking 1927 paper in which he derived both the Einstein $B$ coefficients for induced atomic emission and absorption, and the $A$ coefficients for spontaneous emission, effectively took as his interaction term $e \vec A_t \cdot \dot {\vec X} $, where $e \dot {\vec X} $ is the atomic electric dipole moment operator.\cite{Dirac:1927aa} In his follow up in which he calculated dispersion relations and line widths, he seemed to have produced a truly relativistic Hamiltonian in that he replaced the electron momentum four-vector operators $p_\mu$ by $p_\mu + \frac{e}{c} A_\mu$.\cite{Dirac:1927ab} Yet closer inspection reveals some troubling inconsistancies. He actually interprets $\vec A$ as the transverse vector field operator, while $A_0$ is treated as an external field. 

 In what we shall see constitutes a related development, Jordan and Klein proposed in 1927 a non-relativistic generalization for the electrostatic interaction of charged bosons.\cite{Jordan:1927aa}. They added to the Hamiltonian a Coulomb interaction in which the charge densities were taken to be $e \psi^* \psi $, with the $\psi$ now understood as second-quantized matter field operators. So the interaction took the form $e^2 \int d^3\!x  \int d^3\!x' \psi^*(x) \psi(x) \psi^*(x') \psi(x')/| x - x'| $. Pauli in fact cites this paper, in a letter to Kramers on February 7, 1928, in which he says that he and Heisenberg ``... have together tried to treat the presence of charged particles in an analogous [to the Jordan and Pauli free electromagnetic field] (relativistically invariant) manner. We rely essentially on the results of Jordan and Klein. It seems to work, but itÕs not yet finished; we have difficulties at that point corresponding to Klein and JordanÕs reordering of factors in the energy expression (elimination of the self-energy of the particle).''\footnote{``Inzwischen haben Heisenberg und ich gemainsam auch den Fall des Vorhandenseins geladener Teilchen in analoger (relativistische-invarianter) Weise zu behandeln versucht, wobei wir uns auch wesentlich auf die Resultate von Jordan und Klein st\"utzen. Es scheint in der Tat zu gehen, aber es ist noch nicht alles fertig; wir haben Schwierigkeiten an der Stelle, die der Umstellung der Faktoren im Energieausdruck (Elimination der Wechselwirkingsenergie der Teilchen mit sich selber) bei Klein und Jordan entspricht.''\cite{Hermann:1979aa}, p. 432}
 Two weeks later Pauli actually wrote to Dirac, saying he ``would like to ask [Dirac's] opinion concerning an essential physical difficulty that has come up in the model of Heisenberg and myself. We have been unable to resolve it.)''  I mention this episode here mainly to stress that in February 1928 Heisenberg and Pauli were already committed to the second quantization of a fully interacting matter field, and in their initial attempts at introducing a relativistic interaction Hamiltonian they had already encountered the problem with infinite particle self-energies. This difficulty was not related to the vanishing momentum problem, and it would plague the theory for years to come. It is also of interest that Pauli would address his question to Dirac - who was for decades opposed to second quantization. 

The source of the vanishing momentum problem is gauge invariance, so a short account of the history of this notion will be appropriate here.  It was of course well known at this time that electric and magnetic fields are invariant under the gauge transformations $\delta A_\mu = A_\mu + \xi_{, \mu}$ where $\xi$ is an arbitrary spacetime function. Indeed, in 1929, in one of the most significant papers in 20'th century theoretical physics, Hermann Weyl argued that the electromagnetic potential must exist, with its gauge freedom (so named by him), so as to guarantee the local ray independence of quantum mechanical wave functions.  In particular, the gauge freedom in the potential accompanies the freedom in local phase of wave functions, $\psi \rightarrow \psi e^{ie \xi/\hbar c}$.\cite{Weyl:1929aa} 

Heisenberg's {\it Kunstgriff} of early 1928 was the addition to the electromagnetic Lagrangian of a term $\frac{\epsilon}{2} (A^\mu_{,\mu})^2$.  It appeared in early 1929 in the first of two foundational papers on quantum electrodynamics co-authored by Pauli and Heisenberg.\cite{Heisenberg:1929aa} The idea was to let $\epsilon$ go to zero at the end of all calculations. They point out that without this additional term Gauss' law takes the form of a constraining relation between the electromagnetic 3-momentum ( the electric field $\vec E$), and  the Dirac electron matter field $\psi$ with its conjugate momentum $p_\psi = i \hbar \psi^\dagger$: $\vec \nabla \cdot \vec p = -e \psi^\dagger \psi$. Thus the electric field would possess a longitudinal contribution that would not commute (or anti-commute) with the matter field - even for finite spacelike separations. ``A theory with such non-infinitesimal commutation relations would seem from a practical standpoint to be hopeless, and it would bring with it as well extreme complications in proving relativistic invariance.''\footnote{``Eine Theorie mit solchen nicht infinitesimalen V.-R. durchzuf\"uhren, scheint aber praktisch aussichtslos, zumal der Beweis der relativistischen Invarianz solcher V.-R. mit den gr\"ossten Schwierigkeiten verbunden sein d\"urfte.'' \cite{Heisenberg:1929aa}, p. 30}. They  did recognize, of course, that this modified Lagrangian violated full gauge invariance.  In September, 1929, they offered an alternative (partially) gauge invariant approach.\cite{Heisenberg:1930aa} They proposed to simply set $A_0 = 0$. This choice restricts the gauge descriptor $\xi$ to an arbitrary  time-independent function. Since $A_0$ no longer appeared in their Lagrangian, there was no longer a variation that produced Gauss's law. On the other hand, they recognized that their action was still invariant under the variations $\delta A_a  = \xi_{,a}$ and $\delta \psi = \frac{i e}{\hbar c}\psi  \xi$. Furthermore, they noted that the canonical generator of these variations, $\int d^3\!x C$, was a constant of the motion. Choosing to let this constant of the motion vanish, they recovered Gauss's law: $C = \vec \nabla \cdot \vec p + e \psi^\dagger \psi = 0$. Then, to avoid the problem mentioned above, they required that this constraint be satisfied only as a condition on physically permissible quantum states. A substantial portion of the paper was devoted to proving that despite having singled out the temporal component of the potential for special treatment, the model was never-the-less Lorentz covariant. 

Meanwhile, in the time between the appearance of the first and second paper, Heisenberg and Pauli became aware of an independent approach by Enrico Fermi.\cite{Fermi:1929aa}. Fermi worked from the start with transverse $\vec A$ Fourier coefficients, and simply inserted a Coulomb interaction by hand. In their second paper Heisenberg and Pauli showed that Fermi's model could be derived from a Lagrangian with a gauge-fixing term $-\frac{1}{2} (A^\mu_{, \mu})^2$, with the understanding that the resulting Lorenz condition could only be imposed as a condition on states,  similarly to their treatment of the Coulomb gauge.

Pauli  and Heisenberg were both uneasy about the formal tricks that were undertaken in these various approaches. This is evident both form their correspondence from this period, and their later published recollections.  Gregor Wentzel quotes Pauli as have having remarked, with regard to the proof of Lorentz covariance, ``I forewarn the curious''.\footnote{``Ich warne Neugierige'' \cite{Wentzel:1973aa}, p. 382}  In an interview with Thomas Kuhn in 1963, Heisenberg made the following observations: ``... Also in those papers which Pauli and I wrote on the quantization of fields we saw quite soon that after all it didn't look too well. It is true that for the free light quanta everything could be made to fit, but as soon as interaction came in it didn't look right.''\footnote{AHQM, 2/28/1963, p. 22}  ``... Here, in electrodynamics, it didn't become simple ... For instance, you had to introduce this supplementary condition and you had to make some kind of limiting process -- first introducing an epsilon and at the end you put epsilon to zero. You know, that kind of stuff didn't look right.''\footnote{AHQM, 2/28/1963, p. 23}  ``... Already there it was a bit artificial to do the Lorenz condition without introducing the indefinite metric. Well, finally Pauli and I succeeded in replacing it by some symmetry argument, but again it was a bit funny. You could say that the fourth Maxwell equation is not a rigorous operator equation, itÕs only a supplementary condition to the --, Well, you know. It came into the region of the `Ausrede'.''\footnote{AHQM, 2/28/1963, p. 7}

\section{Pauli and Rosenfeld}

This is where Leon Rosenfeld enters the story. Born in Belgium in 1904, he completed his doctorate in physics at Li\`ege in 1926. There followed several post-doctoral appointments, first in Paris in 1926-27 under the guidance of de Broglie, Brillouin, and Langevin, then in G\"ottingen in 1927-28 under the direction of Max Born. Not incidentally, as Rosenfeld noted in an interview of Oskar Klein that he conducted with John Heilbron in 1963, Rosenfeld and Dirac were together in G\"ottingen in 1927, so their acquaintance dates at least from that time.\footnote{AHQM, 2/28/1963, p. 8} From there Rosenfeld went to work with Pauli in Z\"urich. In Rosenfeld's own words: ``I came to Z\"urich before the summer semester ... I came from G\"ottingen where I was still at the time. I had already corresponded with Bohr, asking him whether I could come to Copenhagen ...  and so I wrote to Pauli then to ask him if he would take me up. He was very friendly and he said: ``With pleasure, because we have just completed a scheme of quantum electrodynamics with Heisenberg; ` dass ist ein Gebiet, dass noch nicht abgebrochen ist.' So he was eager to have people brush up the details and explore the consequences and that is what I did at Z\"urich actually.''\footnote{AHQM, 7/19/1963, p. 5}. Continuing, he remarked that ``É I got provoked by Pauli to tackle this problem of the quantization of gravitation and the gravitation effects of light quanta, which at that time were more interesting. When I explained to Pauli what I wanted to work out, I think it was the Kerr effect or some optical effect, he said ` Well, you may do that, and I am glad beforehand for any result you may find.' That was a way of saying that this was a problem that was not instructive, that any result might come out, whereas at that time, the calculation of the self energy of the light quantum arising from its gravitational field was done with a very definite purpose.''\footnote{AHQM, 7/19/1963, p. 8}  And  ``...Then Pauli told me that he was not at all pleased with longitudinal waves, so he wanted to have them treated another way, which I did, but that was not more enlightening, far from it.''\footnote{AHQM, 7/19/1963, p. 9}  

In September 1929 he completed a paper in which he showed that the longitudinal and scalar electromagnetic potential waves do not couple  to the transverse waves under dynamical evolution, just as Heisenberg and Pauli had hoped.\cite{Rosenfeld:1929aa}
And in September 1930 he submitted a disappointing  demonstration that gravitational interaction produces an infinite photon self-energy.\cite{Rosenfeld:1930ab}  But his principle objective in this period was to investigate the implications of gauge invariance in Hamiltonian dynamics. Referring to the work of Heisenberg and Pauli, he noted in 1963 that `` ... There was this point in their proof in which the invariants of the Hamiltonian seemed to depend on a special structure of the Hamiltonian, and that looked suspicious ... ` Yes, I understand that [said Pauli], but we have not been able to find a mistake in our calculation and we do not understand what this means; we suspect that it must be wrong, but we don't know.' Then the thing came to a crisis through the fact that I tried to make a more general formulation of field quantization É It was a purely abstract scheme which worked in a completely general way with only this complication of accessary conditions, but at any rate, not due to any special structure but only to the existence of invariance with respect to a group. So at that stage I was convinced that there must be a mistake in the original paper ...''\footnote{AHQM, 7/19/1963, p. 5} Indeed there was a mistake. The authors had claimed that a special condition had to be obeyed by the Hamiltonian in order that the theory be Lorentz covariant. Rosenfeld published a note in 1930 in which he showed that this spurious condition arose due to computational error.\cite{Rosenfeld:1930ac} Abraham Pais recounts having been informed by Rosenfeld that ``he [Rosenfeld] regarded his one-and-a-half page article as his best contribution to physics.''\footnote{\cite{Pais:1986aa}, p. 342}

\section{Rosenfeld's constrained Hamiltonian dynamics formalism}

Rosenfeld's 1930 {\it Annalen der Physik} article lays out a Hamiltonian formalism that was intended to apply to all fundamental interactions that were known at the time. It would not only provide the theoretical justification for the gauge fixing techniques employed by Heisenberg and Pauli, but also serve as a first step in the canonical quantization of gravity. The latter effort failed, very likely due to an error in the treatment of general covariance. ( I have found no evidence that Rosenfeld ever recognized the mistake that I will briefly describe below.)  But my objective in this paper is to show that Rosenfeld's treatment of electromagnetic gauge symmetry - and indeed modern generalizations to Yang-Mills type symmetries - is valid and complete.  Then we will inquire why no one seems to have cared. 

I will accompany my description of the Rosenfeld's general theory with his application of the theory to the charged electron field in interaction with the electrodynamic field in flat spacetime. Also, although he formulated his theory in both a c-number and q-number version (making the requisite factor-ordering choices in the q-number version), I will confine this discussion to the classical phase space context. 

Rosenfeld notes in his introduction that identities among configuration and momentum variables arise as a consequence of the invariance of actions under gauge symmetry groups. He mentions that as he  ``...was investigating these relations in the especially instructive example of gravitation theory, Professor Pauli helpfully indicated to me the principles of a simpler and more natural manner of applying the Hamiltionian procedure in the presence of identities. This procedure is not subject to the disadvantages of the earlier methods.''\footnote{``Bei der n\"aheren Untersuchung dieser Verh\"altnisse an Hand des besonders lehrreichen Beispieles der Gravitationstheorie, wurde ich nun von Prof. Pauli auf das Prinzip einer neuen Methode freundlichst hingewiesen, die es in durchaus einfacher und nat\"urlicher Weise gestattet, das Hamiltonsche Verfahren beim Vorhandsein von Identit\"aten auszubilden, ohne den Nachteilen der bisherigen Methoden ausgesetzt zu sein.''\cite{Rosenfeld:1930aa}, p. 114} Unfortunately he does not specify precisely what Pauli had recommended.

Rosenfeld considers Lagrangian densities ${\cal L}$ that are quadratic in first derivatives of dynamical field variables, represented generically by $Q_\alpha$, where  $\alpha$ could be a tensor or internal index. So ${\cal L} = \frac{1}{2} {\cal A}^{\alpha \nu| \beta \mu}(Q) Q_{\alpha,\nu}Q_{\beta, \mu} + \cdots$. The general formalism includes infinitesimal coordinate transformations of the form 
\beq
\delta x^\nu = a^{\nu}_r (x) \xi^r (x) + a^{\nu | \sigma}_r (x) \frac{\partial \xi^r}{\partial x^\sigma}, \label{xtransform}
\eeq
but we will not treat in this paper symmetries of our equations of motion under general coordinate transformations. We will however have symmetry variations of the type 
\beq
\delta Q_\alpha = c_{\alpha r} (x,Q) \xi^r (x) + c^\sigma_{\alpha r} (x,Q) \frac{\partial \xi^r }{\partial x^\sigma},  \label{Qvariation}
\eeq
where the $\xi^r$ are taken to be $r_0$ arbitrary functions of the spacetime coordinates and we assume that the coefficients $c_{\alpha r}$ and $c_{\alpha r}$ do not depend on derivatives of $Q$. If ${\cal L}$ transforms as a scalar density under (\ref{xtransform}) and (\ref{Qvariation}), then the equations of motion will be covariant under this symmetry. The identity 
\beq
\delta {\cal L} + {\cal L} \frac{\partial \delta x^\mu}{\partial x^\mu} \equiv 0, \label{identity}
\eeq
 (and it's generalization to include variations that differ by a total divergence) is the foundation on which the constrained Hamiltonian formalism is built. This identity is, of course, the basis of Noether's famous theorems\cite{Noether:1918aa}, as Rosenfeld duly noted. He is able to derive additional consequences.\footnote{See \cite{Brading:2002aa} for a useful perspective on the relation between Noether's theorems and the conservation of electric charge.}

\begin{changemargin}{.3in}{0in} 
Rosenfeld's flat space electromagnetic Lagrangian is
\beq
{\cal L}_{em} =  -\frac{1}{4} F^{\mu \nu} F_{\mu \nu} + e A_\mu \bar \psi \gamma^\mu \psi +i \hbar c \bar \psi \gamma^\mu \psi_{, \mu} - mc \bar \psi \psi,
\eeq
where $F_{\mu \nu}  := \partial_\mu A_\nu - \partial_\nu A_\mu$ is the electromagnetic tensor and $A_\mu = ( V, -\vec A)$ the electromagnetic 4-potential. (We employ Rosenfeld's metric signature of $-2$.) $\psi$ is the Dirac spinor electron field. This Lagrangian is invariant, i.e., $\delta {\cal L}_{em} \equiv 0$, under the combined transformations
\beq
\delta A_\mu = \xi_{, \mu} =: c^\nu_\mu \xi_{, \nu},
\eeq
and
\beq
 \delta \psi = \frac{ie}{\hbar c} \psi \xi,
 \eeq
where $\xi$ is an arbitrary spacetime function.
\end{changemargin}

Let us now undertake the transition to a canonical phase space description. We note first that the momenta conjugate to the $Q_\alpha$ are 
\beq
p^\alpha = \frac{\partial {\cal L}}{\partial \dot Q_\alpha} = {\cal A}^{\beta \nu|\alpha 0} Q_{\beta, \nu} = {\cal A}^{\beta 0|\alpha 0} \dot Q_{\beta} + {\cal A}^{\beta a|\alpha 0} Q_{\beta, a}. \label{momenta}
\eeq
There are three immediate consequences of the identities (\ref{identity}).
 First, observe that we may write the variation of ${\cal L}$ under the symmetry transformations as
\beq
\delta {\cal L} = \frac{\partial  {\cal L}}{\partial \dot Q_\alpha} \delta \dot Q_\alpha + \cdots
= p^\alpha c^0_{\alpha r} \ddot \xi^r + \cdots \equiv - {\cal L} \frac{\partial \delta x^\mu}{\partial x^\mu},
\eeq
where we have isolated the unique terms containing the second time derivatives of the descriptors $\xi^r$. Thus we deduce the existence of ``primary'' constraints
\beq
p^\alpha c^0_{\alpha r} \equiv 0.
\eeq
The identity holds when the $p^\alpha$ are taken as the function of configuration and velocity variables given by (\ref{momenta}).

Second, the primary constraints give us null vectors of the Legendre matrix $\frac{\partial ^2{\cal L}}{\partial \dot Q_\alpha \partial \dot Q_\beta}$ since
\beq
 \frac{\partial }{ \partial \dot Q_\beta}\left( p^\alpha c^0_{\alpha r} \right) = \frac{\partial  ^2{\cal L}}{\partial \dot Q_\alpha \partial \dot Q_\beta} c^0_{\alpha r} = \frac{1}{2}  {\cal A}^{\beta 0| \alpha 0} c^0_{\alpha r} \equiv 0. \label{null}
\eeq

Third, since this last relation implies that ${\cal A}^{\beta 0| \alpha 0}$ is a singular matrix, (\ref{momenta}) cannot be inverted to express the velocities as unique functions of the momenta.  Indeed, if  $\dot {}^0\!Q_\alpha$ is a  particular solution of (\ref{momenta}), then as a consequence of (\ref{null}), so is $ {}^0\!\dot{Q}_\alpha + \lambda^r c^0_{\alpha r}$, where the $\lambda^r$ are arbitrary spacetime functions.  Furthermore, by diagonalizing the matrix ${\cal A}$ through a similarity transformation  one can construct an explicit particular solution with the property that $r_0$ of the transformed $\dot Q$ vanish and the remainder are unique functions of the momenta. Inserting this expression into ${\cal H}_0 := p^\alpha  {}^0\!\dot{ Q}_\alpha - {\cal L}(Q,  {}^0\!\dot{ Q})$ one obtains an explicit function of the configuration and momentum variables with the property that $ {}^0\!\dot{ Q}_\alpha = \frac{\partial {\cal H}_0}{\partial p^\alpha}$.\footnote{see \cite{Salisbury:2009aa} for details}. Summarizing, Rosenfeld finds that the Hamiltonian is ${\cal H} = {\cal H}_0 + \lambda^r p^\alpha c^0_{\alpha r} $ with the usual equations of motion $\dot Q_\alpha = \frac{\partial {\cal H}}{\partial p^\alpha}$ and $\dot p^\alpha = -\frac{\partial {\cal H}}{\partial Q_\alpha}$. The latter equation of motion follows from the fact that $\lambda^r p^\alpha c^0_{\alpha r}$ does not contribute since the $p^\alpha c^0_{\alpha r}(Q)$ are constrained to vanish.

\begin{changemargin}{.3in}{0in} 
The canonical momenta are $p^\mu =\frac{1}{c} F^{\mu 0}$ and $p_\psi = i \hbar  \psi^\dagger$. Note that since $c^0_\mu = \delta^0_\mu$, the sole primary constraint is $p^\mu c^0_\mu = p^0 \equiv 0$. We find $c p^a = \partial^a A^0 - \partial^0 A^a   =- V_{,a} - \frac{1}{c}\dot A^a = E^a$, where $\vec E$ is the electric field. Solving for $\dot A^a$ we find $\dot A^a  = - c^2 p^a -c V_{,a}$.  Since the Legendre matrix  $\frac{ \partial p^\mu}{\partial \dot A_\nu} =: {\cal A}^{\mu \nu} = \eta^{\mu 0} \eta^{\nu 0} -  \eta^{\mu \nu} \eta^{0 0} $ is already diagonal with ${\cal A}^{0 0} = 0$, we take  as our particular solutions $ {}^0\!\dot{ A}_a  = c^2 p^a + c V_{,a}$ and ${}^0\!\dot{A}_0 = 0$. The resulting Hamiltonian is ${\cal H}_{em} = \frac{1}{2} \left(c^2 \vec p^2 + \vec B^2 \right) + c p^a A_{0, a} -\frac{e}{i \hbar } p_\psi \psi A_0 
-\frac{e}{i \hbar } p_\psi \gamma^0 \gamma^a \psi A_a
- c p_\psi \gamma^0 \gamma^a \psi_{,a}+ \frac{mc}{i \hbar } p_\psi \gamma^0 \psi +  \lambda p^0$. Note that $\dot A_0 = \left\{A_0  , H_{em} \right\} = \lambda$; the time development of $A_0$ is determined through the choice of the arbitrary spacetime function $\lambda$.

\end{changemargin}

Rosenfeld's next step was to construct the phase space generators of infinitesimal symmetry transformations. I will confine my attention here to the case $\delta x^\mu = 0$. It turns out that there is a subtle error in his discussion of general covariance, and I will address this issue elsewhere.\footnote{The problem is that for some variations, conceived as functions of the configuration and velocty variables, there exists no corresponding variation in phase space; the variation is not projectable under the Legendre transformation. This limitation was apparently first noted in print, in 1949,  by Bergmann and Brunings.\cite{Bergmann:1949aa} I will analyze this aspect of Rosenfeld's formalism in a forthcoming paper in {\it Archive for History of Exact Sciences}.} Rosenfeld shows that the generator  ${\cal M} = p^\alpha \delta Q_\alpha = p^\alpha( c_{\alpha r} (x,Q) \xi^r (x) + c^\sigma_{\alpha r} (x,Q) \frac{\partial \xi^r }{\partial x^\sigma} )$ generates the correct symmetry variations of both the configuration and the momentum variables. That we get the correct variation of $Q_\alpha$ is obvious. It is remarkable that we do also get the correct variation of $p^\beta$. The demonstration uses the invariance of the Lagrangian, and proceeds as follows. It is straightforward to show that
\beq
\delta \left( \frac{\partial {\cal L}}{\partial Q_{\alpha, \nu}} \right) -
\frac{\partial}{\partial Q_{\alpha, \nu}}  \delta {\cal L} = - 
\frac{\partial  {\cal L}}{\partial Q_{\beta}}  \frac{\partial }{\partial Q_{\alpha, \nu}} \delta Q_\beta - \frac{\partial  {\cal L}}{\partial Q_{\beta, \mu}}  \frac{\partial }{\partial Q_{\alpha, \nu}} \delta Q_{\beta, \mu}. \label{ident}
\eeq
But $\delta  {\cal L} = 0$, and the first term on the right vanishes due to our assumption (\ref{Qvariation}). Indeed, referring to this equation we find that $ \frac{\partial }{\partial Q_{\alpha, \nu}} \delta Q_{\beta, \mu} = \delta^\nu_\mu \frac{\partial }{\partial Q_{\alpha}} \delta Q_{\beta} $. Thus, taking $\nu = 0$ in (\ref{ident}) we find
\beq
\delta p^\alpha = - p^\beta  \frac{\partial }{\partial Q_{\alpha}} \delta Q_{\beta}.
\eeq
This is precisely the variation of $p^\alpha$ engendered by $\int d^3\!x p^\beta \delta Q_{\beta}$.

Rosenfeld next shows that $\int d^3\!x p^\beta \delta Q_{\beta}$ is a constant of the motion, since under the symmetry transformations
$$
0 \equiv \delta {\cal L} = \frac{\delta  {\cal L}}{\delta Q_\alpha} \delta Q_\alpha
+ \frac{\partial }{\partial x^\mu} \left( \frac{\partial  {\cal L}}{\partial Q_{\alpha, \mu} } \delta Q_\alpha \right), 
$$
where $\frac{\delta  {\cal L}}{\delta Q_\alpha}$ are the Euler-Lagrange equations.
Therefore under suitable boundary conditions at spatial infinity, 
$$
\frac{\partial }{\partial x^0} \int d^3\!x p^\alpha( c_{\alpha r}  \xi^r (x) + c^\sigma_{\alpha r}  \frac{\partial \xi^r }{\partial x^\sigma} ) = \frac{\partial }{\partial x^0} \int d^3\!x \left( (p^\alpha c_{\alpha r} 
-(p^\alpha c^a_{\alpha r})_{,a} )  \xi^r  +  p^\alpha c^0_{\alpha r}  \dot \xi^r \right) = 0.
$$ 
where in the second equality we have performed an integration by parts.

At this stage Rosenfeld makes a remarkable contribution for which he has not received recognition. He argues that since the $\xi^r$ have arbitrary spacetime dependence, one can perform the time derivative and obtain a relation between the necessarily vanishing coefficients of each order of time derivative of $\xi^r$. From
\bea
0 &=&  \int d^3\!x \left( p^\alpha c^0_{\alpha r}  \ddot \xi^r +
 \left( \frac{\partial }{\partial x^0} \left( p^\alpha c^0_{\alpha r}\right)
 +  \left(p^\alpha c_{\alpha r} -(p^\alpha c^a_{\alpha r})_{,a}\right) \right) \dot \xi^r  
 \right. \nonumber \\
 &+& \left.  \left( \frac{\partial }{\partial x^0} \left( p^\alpha c_{\alpha r} -(p^\alpha c^a_{\alpha r})_{,a}  \right) \right)\xi^r
\right),
\eea
he  confirms that $p^\alpha c^0_{\alpha r} = 0$. (Recall that these are the primary constraints.) But then he also finds that the preservation of the primary constraint under time evolution leads to more constraints,
$$
0 = \frac{\partial }{\partial x^0} \left( p^\alpha c^0_{\alpha r} \right) = 
-p^\alpha c_{\alpha r} +(p^\alpha c^a_{\alpha r})_{,a},
$$
and that it must apparently follow from the equations of motion that the generation of further constraints terminates, i.e., 
$$
\frac{\partial }{\partial x^0} \left( p^\alpha c_{\alpha r} -(p^\alpha c^a_{\alpha r})_{,a}  \right)
 = 0.
$$
This discovery has until now been attributed to P. G. Bergmann and J. L. Anderson \cite{Anderson:1951aa}, and it was they who introduced the terminology of primary and secondary constraints, etc.

\begin{changemargin}{.3in}{0in} 
We confirm that 
$ \int d^3\!x {\cal M}_{em} = \int d^3\!x \left( p^\mu \xi_{,\mu} + \frac{ie}{\hbar c} p_\psi   \psi \xi \right) = \int d^3\!x \left( \frac{1}{c}p^0 \dot \xi + \left(  \frac{ie}{\hbar c} p_\psi   \psi  - p^a_{,a} \right) \xi \right)$ generates the correct gauge transformations: $
\delta A_0 = \left\{ A_0,  \int d^3\!x {\cal M}_{em} \right\} = \dot \xi $, 
$ \delta A_a = \xi_{,a}$, $\delta p^0 = \delta p^a = 0$, $\delta \psi =  \frac{ie}{\hbar c}   \psi $, and $\delta p_\psi = - \frac{ie}{\hbar c} p_\psi$.

Let us also calculate $\dot p^0 = \left\{ p^0, H_{em} \right\} =-c p^a_{,a} + \frac{ie}{\hbar } p_\psi   \psi $. This is the required secondary constraint. We of course recognize it as Gauss' Law $\vec \nabla \cdot \vec E = \rho$, with the charge density $\rho = \frac{j^0}{c} = -e \psi^\dagger \psi$.  Let us also check that its time derivative vanishes. For this purpose we require
$$
\dot p^a = \left\{ p^a, H_{em} \right\} = - \nabla^2 A^a + \partial_a \vec \nabla \cdot \vec A 
+ \frac{e}{i \hbar } p_\psi \gamma^0 \gamma^a \psi,
$$
$$
\dot \psi = - c \gamma^0 \gamma^a \psi_{,a} - \frac{e}{i \hbar} \psi A_0 - \frac{e}{i \hbar} \gamma^0 \gamma^a \psi A_a + \frac{mc^2}{i \hbar} \gamma^0 \psi,
$$
and
$$
\dot p_\psi = - c p_{\psi, a} \gamma^0 \gamma^a 
+ \frac{e}{i \hbar} p_\psi A_0 + \frac{e}{i \hbar} p_\psi \gamma^0 \gamma^a  A_a - \frac{mc^2}{i \hbar} p_\psi \gamma^0
$$
The first is of course Maxwell's suitably modified Amp\`ere's Law, $\vec \nabla \times \vec B = \frac{1}{c} j^a + \frac{1}{c} \dot {\vec E}$, where $j^a = e \bar \psi \gamma^a \psi$. The second is the Dirac equation and the third is the Hermitian conjugate of the Dirac equation.  It follows that
$$
\frac{\partial}{\partial t} \left(   p^a_{,a} - \frac{ie}{\hbar c } p_\psi   \psi  \right)
= \frac{e}{i \hbar } \left( p_\psi \gamma^0 \gamma^a \psi \right)_{, a} 
- \frac{i e}{\hbar c} \left( - c p_{\psi, a} \gamma^0 \gamma^a \psi  - cp_psi  \gamma^0 \gamma^a \psi_{,a}  \right) = 0.
$$

\end{changemargin}

\section{The impact of Rosenfeld's work}

Rosenfeld showed in this groundbreaking 1930 paper how to realize the full gauge group of electromagnetism as a canonical transformation group. Thus he provided a group-theoretical justification for all of the gauge choices of Heisenberg, Pauli, and Fock. Yet before he was rediscovered by J. L. Anderson in 1951\footnote{Anderson informed me in July, 2007, that he believed it was he who had discovered Rosenfeld's paper and brought it to the attention of his thesis advisor, P. G. Bergmann}, very few researchers ever cited his work. In fact, even Rosenfeld himself seems not to have thought that the work was significant. He left it out of the volume of his selected works edited by Robert S. Cohen and John J. Stachel.\cite{Cohen:1979aa} In a contemporary overview of quantum electrodynamics published in 1932, he did at least review the general constrained Hamiltonian dynamics formalism.\cite{Rosenfeld:1932aa} But then he reverted to the Coulomb gauge that had been employed by Heisenberg and Pauli without describing in detail how his formalism justified that choice.  In a brief autobiographical sketch he had only this to say about his work in Z\"urich: ``In Z\"urich, Rosenfeld participated in the elaboration of the theory of quantum electrodynamics just started by Pauli and Heisenberg, and he pursued these studies during the following decade; his main contributions being a general method of representation of quantized fields taking explicit account of the symmetry properties of the fields, a general method for constructing the energy-momentum tensor of any field, a discussion of the implications of quantization for the gravitation field, and the proof that a new formulation of quantum electrodynamics proposed by Dirac was not an alternative to the original Heisenberg-Pauli theory, but simply an equivalent representation of the latter (known in later work as the `interaction representation').''\footnote{Niels Bohr Archive, Rosenfeld correspondence 1.8.1971-1974} In a Curriculum Vitae he compiled apparently in 1958, he does not list the 1930 paper as one of his principal works (although curiously the 1932 review article is the first entry on the list).\footnote{Niels Bohr Archive, 10/11 Copenhagen, 7 Personnelle}. 

Perhaps the clearest indication of Rosenfeld's attitude toward his own invention appears in a review article he wrote in Danish in 1935. He writes ``As Heisenberg and Pauli (6) first showed, the quantum theory of radiation can,  by letting the constraint condition $\vec \nabla \cdot \vec E = 0$ fall,  be directly extended to an invariant and logically closed Òquantum electrodynamicsÓ, which summarizes the quantized electromagnetic field in the most general sense and the matter field associated with the point model as well as their interaction. The proof of the invariance of the field equations and the quantum conditions in their original form demands rather complicated considerations (8)\cite{Rosenfeld:1932aa}, but later Dirac found (9) a very beautiful presentation of the theory, in which its invariance appears directly.''\footnote{``Som Heisenberg og Pauli (6) f\o rst paaviste, kan Kvanteteorien for Straalingen, ved at lade den indskr¾nkede Betingelse  falde, umiddelbart udbygges til en invariant og logisk afsluttet ÓKvanteelektrodynamikÓ, der sammenfatter baade det kvantiserede elektromagnetiske Felt i dets almindeligste Forstand og det til Punktmodellen svarende Materiefelt samt deres Vekselvirkning.
Beviset for Feltligningernes og Kvantebetingelsernes Invarians i disses oprindelige Form kr\ae ver temmelig komplicerede Betragtninger (8), men senere fandt Dirac (9) en meget smuk Fremstillingsmaade af Teorien, hvori dens Invarians umiddelbart tr\ae der frem.
\cite{Rosenfeld:1935aa}, p. 114. I thank Anja Skaar Jacobsen for this translation.} Rosenfeld seems to be signaling here a disenchantment with second quantization, a feeling that was widespread among theorists of this time and in fact into the late 1940's. Besides the fact that Lorentz covariance was not manifest, the theory was beset with infinities. Already in his 1932 review Rosenfeld wrote ``... we have developed the principal consequences of the formalism of quantization applied to the material and electromagnetic fields, and we have seen that in both cases we end up with complete failure.''\footnote{``...nous avons d\'evelopp\'e les principales cons\'equences du formalisme de la quantification appliqu\'e au champ mat\'eriel et au champ \'electromagn\'etique, et nous avons vu que, dans les deux cas, nous aboutissons \`a un \'echec complet.''\cite{Rosenfeld:1932aa}, p. 86}

Interestingly, one of the few authors who did cite Rosenfeld in the 1930's was Oskar Klein.\footnote{\cite{Klein:1997ab}, p. 163} And he did so in the paper that is recognized as the forerunner of modern non-Abelian gauge theory. It is ironic that it is in a 1955 letter from Pauli to this same Klein that we find Pauli's own judgement of Rosenfeld's work. He writes ``I would like to bring to your attention the work by Rosenfeld in 1930. He was known here at the time as the `man who quantised the VierbeinÕ (sounds like the title of a GrimmÕs fairy tale doesn't it?) See part II of his work where the Vierbein appears.  Much importance was given at that time to the identities among the p's and q's (that is the canonically conjugate fields) that arise from the existance of the group of general coordinate transformations. I still remember that I was not happy with every aspect of his work since he had to introduce certain additional assumptions that no one was satisfied with.''\footnote{``Gerne m\"ochte ich Dich in dieser Verbindung auf die lange Arbeit von Rosenfeld, Annalen der Physik (4), 5, 113, 1930 aufmerksam machen. Er hat sie seinerzeit bei mir in Z\"urich gemacht und hiess hier dementsprechend `der Mann, der das Vierbein quantelt' (klingt wie der Titel eines Grimmschen M\"archens, nicht?). - Siehe dazu Teil II seiner Arbeit, wo das `Vierbein' daran kommt. Auf die Identit\"aten zwischen den `p' und `q' - d.h. kanonisch konjugierten Feldern - die eben aus der Existenz der Gruppe der Allgemeinen Relativit\"tstheorie (Koordinaten-Transformationen mit 4 willk\"urlichen Funktionen) entspringen, wurde damals besonderer Wert gelegt. Ich erinnere mich noch, dass Rosenfelds Arbeit nicht in jeder Hinsicht befriedigend war, da er gewisse zus\"atzliche Bedingungen einf\"uhren musste, die niemand richtig verstehen konnte.''\cite{Meyenn:2001aa}, p. 64}

I will conclude this section with the observation that Rosenfeld had all of the tools at his disposal to construct the fields that satisfy the Coulomb gauge condition and are invariant under the full gauge symmetry group. He could then perhaps have convinced more researchers of the utility of his formalism. The modern canonical approach can be found, for example, in the first volume of Weinberg's {\it The Quantum Theory of Fields}.\cite{Weinberg:1995aa} There is however yet one important element missing from even this approach. Weinberg makes use of Dirac brackets to construct variables that have a vanishing bracket algebra with the constraints and gauge conditions. He does not recognize that these brackets are nothing other than the ordinary Poisson brackets of invariant functionals of the fields, and these functionals can be constructed by carrying out an appropriate canonical gauge transformation on the field variables. The idea is that one can determine the field-variable-dependent gauge transformation that transforms the fields in an arbitrary gauge to the fields that satisfy the gauge conditions. Then one can employ this gauge transformation to transform all of the fields. The resulting transformed fields are manifiestly gauge invariant functionals of the original fields.\footnote{I was taught this technique by A. P. Balachandran in 1975. I do imagine that it was known to others then, and very likely earlier. Indeed, Balachandran showed me a proof at that time that the invariants constructed in this way in non-Abelian gauge theories do satisfy the Dirac algebra. A general proof that applies also to generally covariant theories can be found in \cite{Pons:2009aa}} I give the construction in the appendix. Here let me simply cite the results. The gauge fixing conditions are $A^a_{,a} = 0$ and $c p^a_{,a} + \nabla^2 A_0 =0$. The invariant functionals, represented with a ``hat'', are $\hat A^a(\vec x) = A^a (\vec x) +  \frac{\partial }{\partial x^a} \left( \frac{1}{4 \pi} \int d^3\!y A^b_{,b} (\vec y)   \frac{1}{\left| \vec x - \vec y \right|} \right)$ (not surprisingly, the transverse vector potential!),  $\hat A_0(\vec x) =  \frac{c}{4 \pi} \int d^3\!y  p^a_{,a} (\vec y) \frac{1}{\left| \vec x -\vec y \right|} = \frac{i e}{4 \pi\hbar } \int d^3\!y  p_\psi (\vec y) \psi (\vec y) \frac{1}{\left| \vec x -\vec y \right|}  $, $\hat p^a(\vec x) = p^a(\vec x)$, and $\hat \psi (\vec x)= \psi (\vec x) \exp\left(\frac{e}{4 \pi i \hbar } \int d^3\!y \frac{A^a_{,a}(\vec y)}{ \left| \vec x - \vec y \right|} \right)$.These invariant expressions are to be substituted for the corresponding non-gauge-invariant analogues in the Hamiltonian.  And the constraints are to be taken as operator equalities. It is perhaps surprising that the $\hat A_0$ that multiplies the Gaussian constraint disappears. But it reappears in the $ {\vec p}^2$ term since $ p^a =  p^a_t +  p^a_l$ where the longitudinal part of $ p^a$ is $ p^a_l = \frac{\partial}{\partial x^a} \nabla^{-2} p^a_{,a} = -\frac{1}{c} \frac{\partial}{\partial x^a} A_0$. Therefore $\int d^3\!x ( \vec p^2) = \int d^3\!x ( \vec p_t^2 + \vec p_l^2)$ and $  \int d^3\!x  \vec p_l^2  = - \int d^3\!x  \vec p_l^a \hat A_{0, a} = \int d^3\!x  \vec p^a_{l ,a} \hat A_0  = \frac{e^2}{4 \pi \hbar^2 c^c}\int d^3\!x \int d^3\!y \frac{p_\psi(\vec x)  \psi(\vec x) p_\psi(\vec y)  \psi(\vec y)}{\left| \vec x - \vec y \right|} $. This expression is the static Coulomb energy.\footnote{see for example \cite{Schiff:1955aa} for the conventional derivation of this Hamiltonian in the Coulomb gauge.}

\section{Dirac and Rosenfeld}

It is a curious fact that Dirac was already in 1932 familiar with Rosenfeld's work on constrained Hamiltonian dynamics. Rosenfeld had been in correspondence with Dirac concerning Rosenfeld's demonstration of the equivalence of Dirac's formalism to that of Heisenberg and Pauli. Rosenfeld sent him a draft on April 30, 1932, prior to publication, writing that ``I enclose a note about your new theory, which is clearly not meant `um zu kritisieren' but `nur zu lernen'... ''\footnote{Churchill College Archive, DRAC 3} On May 6, 1932 Dirac wrote ``Thank you very much for the paper you sent me. I found it very interesting. The connection which you give between my new theory and the Heisenberg - Pauli theory is, of course, quite general and holds for any kind of field (not simply the Maxwell kind) in any number of dimensions. This is a very satisfactory state of affairs. In the same letter Dirac posed a question to Rosenfeld on the validity of the Heisenberg-Pauli demonstration of Lorentz covariance.\footnote{Niels Bohr Archive, Rosenfeld Papers}. Responding to Dirac on May 10, Rosenfeld suggested that Dirac ``examine the general invariance proof which I give in my paper of the 'Annales de l'Institut Poincar\'e', or in a more elaborate form in my paper of the 'Annalen der Physik' \underline{5}, 113, 1930. (I sent you reprints of both.)... ''\footnote{Churchill College Archive, DRAC 3}  Dirac followed up on May 16: ``I have been studying your papers, but have had some trouble in understanding the significance of your $\lambda$'s. What exactly is meant by the statement that they are arbitrary?''\footnote{Niels Bohr Archive, Rosenfeld Papers} Rosenfeld responds on May 21: ``... As to the $\lambda$'s, they enter as arbitrary or undetermined coefficients (depending on coordinates) in the general expression of the  $\dot Q$     in terms of the Q's and P's. In equation (111) the hamiltonian should be the same as that of Heisenberg-Pauli (as stated there), so that the substitution of the P's in terms of the $\dot Q$'s      in them will lead to identities, and this implies no restriction for $\lambda$. But the purpose of the $\lambda$ - method is not to get a more general scheme than Heisenberg and Pauli, but to give an alternative proof of the invariance of this scheme for the whole gauge and (general) relativity group.''\footnote{Churchill College Archive, DRAC 3} It is perhaps not impertinent to note that beginning with Dirac's first paper on constrained Hamiltonian dynamics in 1950\cite{Dirac:1950aa} he uses the same symbol $\lambda$ to represent the aribrary functions that appear in his formalism.

\section{Conclusion}

In this paper I have first attempted to depict the context in which Rosenfeld offered his own contributions in the developing formalism of quantum field theory. Then I have offered a detailed overview of his invention of constrained Hamiltonian dynamics, with special attention devoted to the implementation of this theory in quantum electromagnetism. It remains a mystery why his contribution went practically unnoticed for two decades. But I have suggested a plausible explanation. It was too early. There were too many difficulties with the second quantization program. Gradually the Dirac program prevailed, in spite of its initial failure in fully accounting for negative energy states. The primary dividend was it's manifest Lorentz and gauge covariance.  It is ironic indeed that Rosenfeld himself contributed to the prevalence of Dirac's theory by demonstrating that his much less complicated formalism was equivalent to the quantized matter field program. Yet - one must still wonder how a physicist of Rosenfeld's caliber, who famously advised his younger colleagues to push for recognition of their achievements\footnote{``According to Rosenfeld, it was the responsibility of the pioneer to bear his ideas to triumph and make sure that they would be accepted by the scientific community, and also more broadly. If he did not succeed, for whatever reasons, he would be reduced to a forerunner in history and the idea would fall into oblivion until it was rediscovered.''\cite{Jacobsen:2007aa}, p. 31. Jacobsen discusses the relation between this dictum, expressed in \cite{Rosenfeld:1938aa}, and Rosenfeld's Marxist philosophy.}, could have failed to do precisley that when gauge theory came into its own beginning in the 1950's.

\section*{Acknowledgements} 

I would like to thank J\"urgen Renn and the Max Planck Institute for the History of Science for providing a stimulating environment  and generous support while most of this work was undertaken. Thanks also for countless fruitful discussions with the members of the History of Quantum Mechanics Study Group. I would especially like to thank Anja Skaar Jacobsen for her insights and Felicity Pors for her invaluable assistance at the Niels Bohr Archive. Thanks also to Kurt Sundermeyer and Josep Pons for their critical comments, and Sandra Marsh at the Churchill Archives Centre for her assistance.
  
\section*{Appendix}

In this appendix I will show how the classical canonical transformation group can be employed to construct invariant functionals corresponding to a gauge choice. Let us represent the primary constraint by $C^1 := \frac{1}{c}p^0 = 0$, and the secondary constraint by $C^2 =-p^a_{,a} - \frac{ e}{i \hbar c} p_\psi \psi = 0$. 
The generator of infinitesimal gauge transformations is
\beq
M(\xi) =\int d^3\!x \left(\frac{1}{c} p^0 \dot \xi -\left( p^a_{,a} + \frac{e}{i \hbar c} p_\psi \psi \right)\xi \right),
\eeq
where the descriptors $\xi$ are understood to be arbitrary infinitesimal spacetime functions. 

We will construct invariant functionals that satisfy the Coulomb gauge condition $\chi_2 := A^a_{,a} = 0$. Preservation of this condition under time evolution results in an additional condition; we require that 
\beq
0 = \left\{A^a_{,a}, H_{em} \right\} = -c^2 p^a_{,a} - c \nabla^2 A_0 = :  \chi_1.
\eeq

We wish to determine the finite gauge transformation that will transform arbitrary solutions of the equations of motion to solutions that satisfy the gauge conditions.  Our infinitesimal gauge generator for infinitesimal $\xi$ is
\beq
M( \xi) =\int d^3\!x \left( C^1 \dot \xi + C^2 \xi \right).
\eeq
 The finite generator for finite $ \xi$ is
\beq
\exp \left( \left\{ - , M( \xi) \right\} \right) :=  1 + \left\{ - , M(\xi )\right\} +
\frac{1}{2} \left\{ \left\{- , M(\xi)\right\},
M(\xi) \right\} + \cdots
\eeq
Let us first find $ \xi(A)$ such that the gauge transformed $\chi_2$ is zero, i.e.,
\beq
0 = \exp \left( \left\{ - , M( \xi) \right\} \right) \chi_2 = \chi_2 + \left\{ \chi_2 , M( \xi)\right\}
= \chi_2 - \nabla^2 \xi, 
\eeq
where we recognize that since $\left\{ \chi_2 , M( \xi)\right\}$ does not depend on the canonical variables, there are no contributions from the nested Poisson brackets.  Thus the required descriptor is $ \xi = \nabla^{-2} A^a_{,a} = \nabla^{-2} \chi_2$. Let us check to see whether this descriptor will yield the correct gauge transformed $\chi_1$. We have
\bea
 \exp \left( \left\{ - , M( \xi) \right\} \right) \chi_1 &=& \chi_1 + \left\{ \chi_1 , M(\xi)\right\}
= \chi_1 + \nabla^2 \dot{ \bar \xi} \nonumber \\
&=& -c^2 p^a_{,a} - c \nabla^2 A_0 - \dot A^a_{,a} = 0 ,
\eea
where in the last line we substituted $ \xi = \nabla^{-1} A^a_{,a}$ and used the equation of motion $\dot A^a = -c^2 p^a - c A_{0,a}$. Note that therefore $\dot \xi = \nabla^{-2} \chi_2$. We find therefore that the Coulomb gauge is a legitimate gauge.

Now we have expressions for the gauge invariant fields $\hat \Phi $ associated with all canonical dynamical variables $\Phi$, namely
\bea 
\hat \Phi (\vec x) &=& \Phi(\vec x) + \int d^3\!y \nabla^{-2}\chi_i (\vec y) \{ \Phi(\vec x) , \,
 C^i \} \nonumber \\
 & + &  \int d^3\!y  \int d^3\!z \nabla^{-2} \chi_i (\vec y) \nabla^{-2}\chi_j(\vec z) \{\{ \Phi(\vec x) , \,
 C^i (\vec y)\}, \,C^j(\vec z) \} + \cdots  \label{calIphi}
 \eea
 
 Let us consider first the invariant associated with $A_0$. For this purpose we must calculate
 \bea
&& \int d^3\!y  \nabla^{-2} \chi_i (\vec y) \{ A_0(\vec x) , \,
C^i \} =\frac{1}{c} \int d^3\!y  \nabla^{-2} \chi_1 (\vec y) \{ A_0(\vec x) , \,
 p^0 (\vec y) \}  \nonumber \\
 &=& - \nabla^{-2} \left( c p^a_{,a}(\vec x) +  \nabla^2 A_0 (\vec x) \right) \nonumber \\
 &=&  \frac{c}{4 \pi} \int d^3\!y  p^a_{,a} (\vec y) \frac{1}{\left| \vec x -\vec y \right|}
 -   A_0(\vec x)
 \eea
 Notice that the Poisson brackets $\left\{  A_0  ,  C^i   \right\}$ are independent of the canonical variables, so all the remaining nested brackets in $\hat A_0$  vanish. We deduce therefore that the invariant associated with $A_0$ is
\beq
\hat A_0(\vec x) = A_0  (\vec x) +\frac{c}{4 \pi} \int d^3\!y  p^a_{,a} (\vec y) \frac{1}{\left| \vec x -\vec y \right|} -  A_0  (\vec x) = \frac{c}{4 \pi} \int d^3\!y  p^a_{,a} (\vec y) \frac{1}{\left| \vec x -\vec y \right|} 
\eeq
 
 Next we find the invariant associated with $A^a$. For this purpose we need
 \bea
 && \int d^3\!y \nabla^{-2} \chi_i (\vec y) \{ A^a(\vec x) , \,
 C^i \} = \int d^3\!y \nabla^{-2}\chi_2 (\vec y) \{ -A_a(\vec x) , \,
- p^b_{,b} (\vec y)  \}  \nonumber \\
&=&- \left(  \nabla^{-2}\chi_2 (\vec x) \right)_{, a} = \left( \frac{1}{4 \pi} \int d^3\!y \frac{A^b_{,b}(\vec y)}{\left| \vec x -\vec y \right|} \right)_{,a}
\eea
Once again, higher order nested brackets do not contribute  and we find
\beq
\hat A^a(\vec x) = A^a (\vec x) +  \frac{\partial }{\partial x^a} \left( \frac{1}{4 \pi} \int d^3\!y A^b_{,b} (\vec y)   \frac{1}{\left| \vec x - \vec y \right|} \right).
\eeq
This is the transverse field, i.e.,
$$
\frac{\partial }{\partial x^a} \hat A^a(\vec x) = 0.
$$
Similarly, we find that
\beq
\hat \psi = \psi -\frac{e}{i \hbar c} \psi \chi_2 + \left(\frac{e}{i \hbar c} \right)^2 \chi^2 \psi + \cdots
= \psi e^{\frac{e}{i \hbar c} \nabla^{-2} A^a_{,a} }.
\eeq
The momenta $p^a$ are of course themselves invariant.

 \bibliographystyle{elsarticle-num}
\bibliography{historyqm}

  \end{document}